\documentclass[a4paper]{jpconf}
\usepackage{graphicx}

\begin{document}
\title{Physical Unclonable Functions using speckle patterns of perfect optical vortices}
\author{Patnala Vanitha,$^{1,*}$ Bhargavi Manupati,$^{1}$, Inbarasan Muniraj, $^{1}$ Satish Anamalamudi, $^{1}$  Salla Gangi Reddy,$^{1,**}$ and R. P. Singh$^2$}

\address{$^1$ SRM University-AP, Amaravati, India - 522240. \\
$^2$Physical Research laboratory, Navarangpura, Ahmedabad, India-380009.
\\
$^*$vanitha\_patnala@srmap.edu.in. \\
$^**$Corresponding author: gangireddy.s@srmap.edu.in \\}
\date{\today}

\begin{abstract}

Encryption techniques demonstrate a great deal of security when implemented in an optical system (such as holography) due to the inherent physical properties of light and the precision it demands. However, such systems have shown to be vulnerable during digital implementations under various crypt-analysis attacks. One of the primary reasons for this is the predictable nature of the security keys (i.e., simulated random keys) used in the encryption process. To alleviate, in this work, we are presenting a Physically Unclonable Functions (PUFs) for producing a robust security key for digital encryption systems. To note, a correlation function of the scattered perfect optical vortex (POV) beams is utilized to generate the encryption keys. To the best of our knowledge, this is the first report on properly utilizing the scattered POV in optical encryption system. To validate the generated key, one of the standard optical encryption systems i.e., Double Random Phase Encoding, is opted. Experimental and simulation results validate that the proposed key generation method is an effective alternative to the digital keys. 

\end{abstract}

\section{Introduction}
Ever-increasing demand for the Internet of Things (IoT) based devices mandates the voluminous data transfer over the communication channels. In this context, securing private data while authenticating authorized third-party users to access the sensitive (personal) information becomes mandatory. Several mathematics-based security approaches (cryptography) have been demonstrated to secure the systems and networks from malicious attacks. Thus, cryptographic algorithms play a vital role in today’s digitized and datafied era. In general, information that needs to be sent from sender end, is encrypted (i.e., input data is converted into an unreadable format) using secret keys. At the receiver end, by using appropriate keys encoded information can be retrieved and this process is known as decryption. It is known that, depends on the cryptography algorithm opted, keys for both the encryption and decryption can be same or different \cite{ref1, ref2}. Owing to this capability, cryptographic algorithms are widely used in various fields, for instance, securing bank data, healthcare, social media, and military communication, etc. Over the years, advancements of high-performance computers made the cryptographic methods vulnerable. To improvise the security, for the first time, Miller introduced the One Time Pad (OTP) technique which basically uses a perfect random key to ensure the security \cite{ref3}. 

In a different context, Optical Signal Processing (OSP) based encryption methods shown to be offering great deal of security with lower computational complexity when compared to the digital encryption approaches e.g., Advanced Encryption Standard (AES), Data Encryption Standard (DES), and Rivest–Shamir–Adleman (RSA) to name a few. Double Random Phase Encoding (DRPE) is one of the earliest optical information security methods that demonstrated remarkable advantages, i.e., extended degrees of freedom, system flexibility, multi-dimensional capabilities, and high encryption density, to name a few \cite{drpe_base, inba_book}. In general, optical systems also exhibit an inherent parallel-processing capability i.e., operating independently on the incident information without the need to sequentially process data \cite{osp_book}. In addition, DRPE encrypts both the real and phase information independently and can be extended to encrypt information based on amplitude, polarization, and wavelength etc \cite{drpereview}. Since the first inception of the classical DRPE i.e., Fourier transform based encryption, several variations have also been proposed that includes Fresnel transform (FST)\cite{fst}, Fractional Fourier transform (FrFT)\cite{frft}, Hartley Transform (HT)\cite{ht}, and Linear Canonical Transform (LCT)\cite{qps}. In addition to this, some new types of encryption systems such as compressive sensing based encryption \cite{csdrpe}, 4D light field based microscopic encryption \cite{micro}, and photons-counting imaging based optical encryption \cite{pcidrpe} have also been demonstrated. 

In general, the robustness of a crypto system is relies on the secret key that is being used and the randomness of ciphertext it generates. In this context, the linearity of DRPE mechanism have been exploited to show the vulnerability of DRPE scheme against chosen-plaintext and known-plaintext attacks \cite{attack}. For this purpose, physical one-way functions have been introduced for cryptographic systems that can be (practically) realized using the scattering of light beams \cite{pappu}. We note, these functions can be generated via physical processes which are impossible to duplicate (due to the unique manufacturing process of materials such as silicon chip or ground glass) and has no compact mathematical representation. Owing to these intrinsic randomness, this  approach was shown to be a robust alternate to the standard crypto-systems \cite{puf1, puf5}. These functions are also named as Physical Unclonable functions (PUFs) and can be used for data authentication. Some of the advantages of PUFs include (1) low cost per piece (2) high output complexity (3) difficult to replicate, and (4) high security against attacks \cite{puf2}. To note, the scattering of light beams results in a highly complex and random output i.e., speckles. Their distribution and size are very sensitive to the input laser parameters such as beam width, angle of incidence as well as on the scattering surface and can act as optical PUFs \cite{puf3, puf4, goodman}. To the best of our knowledge, this is the first report on using PUFs data in optical encryption scheme.

In this work, we propose to generate an encryption key by taking a correlation function (beyond optical PUFs) between two speckle patterns obtained after scattering the POV beams through a ground glass plate and demonstrate encryption using DRPE. Rest of the paper is organized as follows: Section 2 describes the basics of DRPE scheme. In Section 3, we discuss the methodology used to generate the proposed encryption key and the experimental and simulation results are provided in Section 4. Finally, Section 5 concludes the paper. 

\section{Double Random Phase Encoding}

\begin{center}
   \begin{figure}[htb]
   \includegraphics[width=14.0cm]{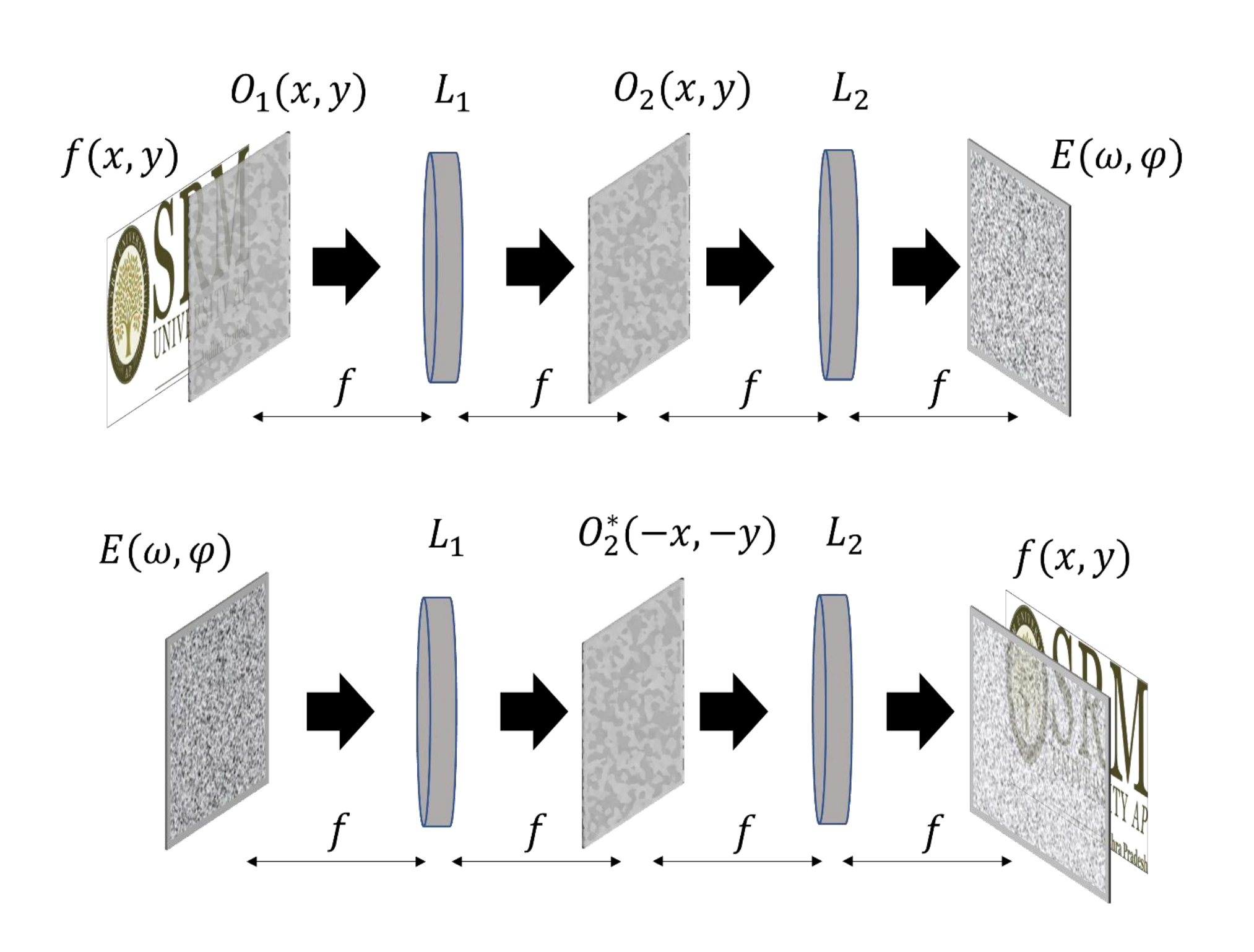}
    \caption{\textit{(Color online) The schematic for DRPE or 4f optical encryption system. Here, $O_1(.)$ and $O_2(.)$ denotes RPMs placed at the spatial and the Fourier plane. $L$ performs optical Fourier transform.}}
    \label{DPRE}
   \end{figure}
   \end{center}

In general, DRPE can be classified based on $(1)$ Amplitude Encoding (AE), and $(2)$ Phase Encoding (PE). We are discussing AE-DRPE in this work and its schematic diagram is given in Fig.\ref{DPRE}). A function \textit{f(x,y)} represents a 2D image which is to be encrypted. In this scheme, two Random Phase Masks (RPMs), $O_1$ and $O_2$ are considered. A collimated input light field passes through a first random phase mask  $O_1(x,y)$, then the product is Fourier Transformed $(FT_1)$. The resultant Fourier spectrum is multiplied with the second phase mask $O_2 (x,y)$, and an inverse Fourier transform is applied $(FT_2)$. To note, RPMs scramble the input light phase by $exp\lbrace j2\pi n_1(x,y)\rbrace$ and $exp\lbrace j2\pi n_2(x,y)\rbrace$ in which $n_1(x,y)$ and $n_2(x,y)$ are considered as secret keys that are generated using scattered POV. To note, the secret keys are statistically independent but uniformly distributed in [0,1]. Mathematically, the encrypted (output) image $E(\omega,\varphi)$ can be expressed as \cite{inba_book}:
\begin{eqnarray}
E(\omega,\varphi)=FT_2\lbrace FT_1\lbrace f(x,y)\times O_1(x,y)\rbrace \times O_2(x,y)\rbrace 
\end{eqnarray}

As can be seen, the resultant encrypted image, $E(\omega,\varphi)$, resembles a speckle image therefore it does not disclose any of the input information. Nevertheless, to note, the input data is just scrambled and not lost. It is therefore possible to reverse this process and get the original image back. The process of reversing an encrypted data into a readable one is known as decryption, and it is exactly a reverse process of encryption. Thus, by employing inverse keys i.e., $O_1^*(-x,-y)$ and $O_2^*(-x,-y)$ and input image \textit{f(x,y)} can be retrieved without a loss. To record a complex encrypted data an optical holographic imaging setup is preferred \cite{pcidrpe,inba_book}.

\section{Generation of encryption keys}

\begin{center}
   \begin{figure}[htb]
   \includegraphics[width=12.0cm]{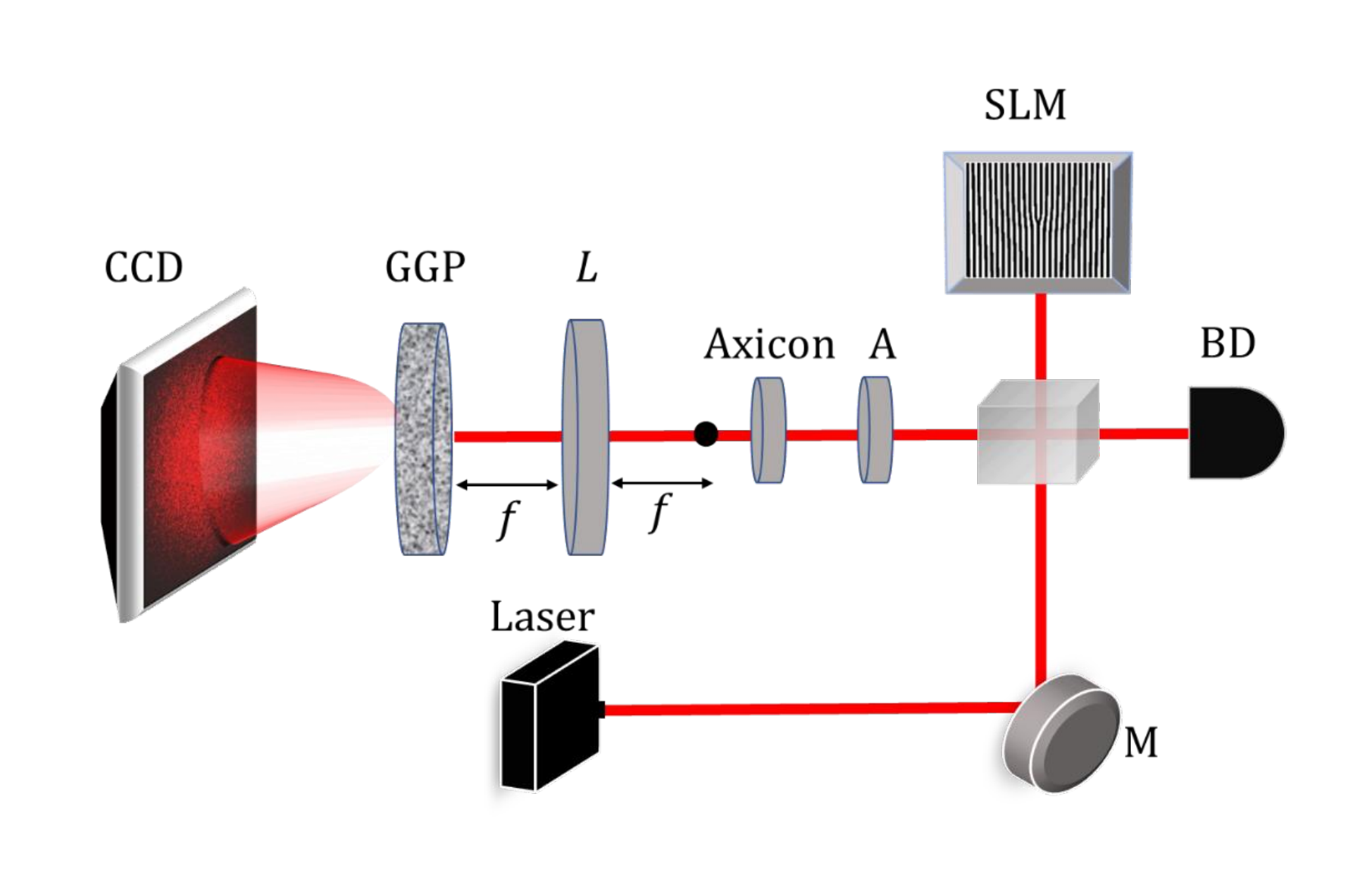}
    \caption{\textit{(Color online) Experimental setup for the generation of perfect optical vortex (POV) beams and the corresponding speckle patterns. Where M-Mirror; BD-Beam Dumper; BS-Beam Splitter; SLM-Spatial Light Modulator; A-Aperture; L-Lens and f-focal length of L. A black dot at the back focal plane of Axicon represents the formation of Bessel-Gaussian beam.}}
    \label{fig:expt}
   \end{figure}
   \end{center}

In this work, the encrypted keys are generated using a correlation function obtained from two scattered POV light beams i.e., speckles. The experimental set up for generating POV beams, and the corresponding speckle patterns is shown in Fig. \ref{fig:expt}. 

A He-Ne laser having a wavelength of 632 nm has been used to generate the vortex beam by combining the illuminated light field with a computer-generated hologram, which is displayed on spatial light modulator (SLM), via a beam splitter (BS). This combined light beam is then propagated through an Axicon of apex angle 178° to convert optical vortex beams to Bessel-Gaussian (BG) beam. To note, the formation of BG beams happens at 12.5cm from axicon (shown as a black colour dot in the optical path in Fig. \ref{fig:expt}) but that BG beam further travels to 60cm (focal length of the lens) to reach the Fourier lens (L). POV beam is generated at the back focal plane in where we have placed a GGP (DG-10-600, from Thorlabs) to scatter the POV beam. Thus, the speckle images are produced and recorded using a charged-coupled device (CCD camera ) (FLIR, pixel size of 3.45 $\mu$m).  
   
To note, after recording the speckle patterns, we have determined the correlation function in MATLAB software and the same is used as an encrypted key $n_1(x,y)$ and $n_2(x,y)$ in the DRPE scheme. The mathematical description for the correlation is as follows:    

The field distribution of a POV beam with a thin annular ring of order $m$, is \cite{ostro}
\begin{equation}
E(\rho,\theta)= \delta\left(\rho-\rho_0\right)\ e^{im\theta}
\label{field}
\end{equation}
where $\rho_0$ is the radius of the POV beam, and $\delta$ represents the Dirac delta function.

The scattering of POV beams through a ground glass plate (GGP) can be described with random phase function $e^{i\Phi}$, where $\Phi$ varies randomly from 0-2$\pi$. The scattered light filed is given by \cite{goodman, dainty, gangi}
\begin{equation}
U(\rho,\theta)= \delta\left(\rho-\rho_0\right)\ e^{im\theta} e^{i\Phi}
\label{Sfield}
\end{equation}

Now, the  mutual coherence function between the two scattered POV fields of same order is defined as  $\Gamma\left(r_1,\theta_1,z \right)=<U_1(r_1,\theta_1,z)U_1^*(r_1,\theta_1,z)>$ where $< >$ denotes the ensemble average. After solving the above using Fresnel's diffraction integral, we get \cite{acevedo}

\begin{eqnarray}
\Gamma_{12}\left( \Delta r\right)=\frac{e^{\frac{ik}{2z}\left( r_1^2-r_2^2\right) } }{\lambda^2 z^2}\int\int  U_1(\rho,\theta)U_1^*(\rho,\theta)  \nonumber  \\
e^{-\frac{ik}{z}\left( \rho\Delta rcos(\varphi_s-\theta)
\right)} \rho d\rho d\theta 
\label{Correlation}
\end{eqnarray} 
  
where
\begin{eqnarray}
\Delta rcos(\varphi_s-\theta)&=&\left[\left( r_1 cos\left( \varphi_1\right) -r_2 cos\left( \varphi_2\right)\right)cos\theta\right]             \nonumber  \\
& + &  \left[\left( r_1 sin\left( \varphi_1\right) -r_2 sin\left( \varphi_2\right)\right)sin\theta\right]
\end{eqnarray} 
and $\Delta r^2=r_1^2+r_2^2-2r_1r_2 cos\left(\varphi_2-\varphi_1\right) $.

with the help of Anger-Jacobi Identity along with the integral properties of Dirac - delta function \cite{ryzhik}, we get the auto-correlation function as: 

\begin{eqnarray}
\Gamma_{12}\left( \Delta r\right)&=&\frac{2\pi\rho_0 e^{\frac{ik}{2z}\left( r_1^2-r_2^2\right) } }{\lambda^2 z^2 }  J_{0}\left( \frac{k\rho_0}{z}\Delta r\right).  
\label{output}
\end{eqnarray}

The normalized intensity distribution of the auto-correlation function can be evaluated in terms of time averaged intensity $I_0$ as \cite{vanitha}
\begin{equation}
I\left( \Delta r\right)={I_0}^2 \left(1+ J_0^2\left( \frac{k\rho_0}{z}\Delta r\right)\right) 
\end{equation}

\begin{center}
\begin{figure}[htb]
   \includegraphics[width=12.5cm]{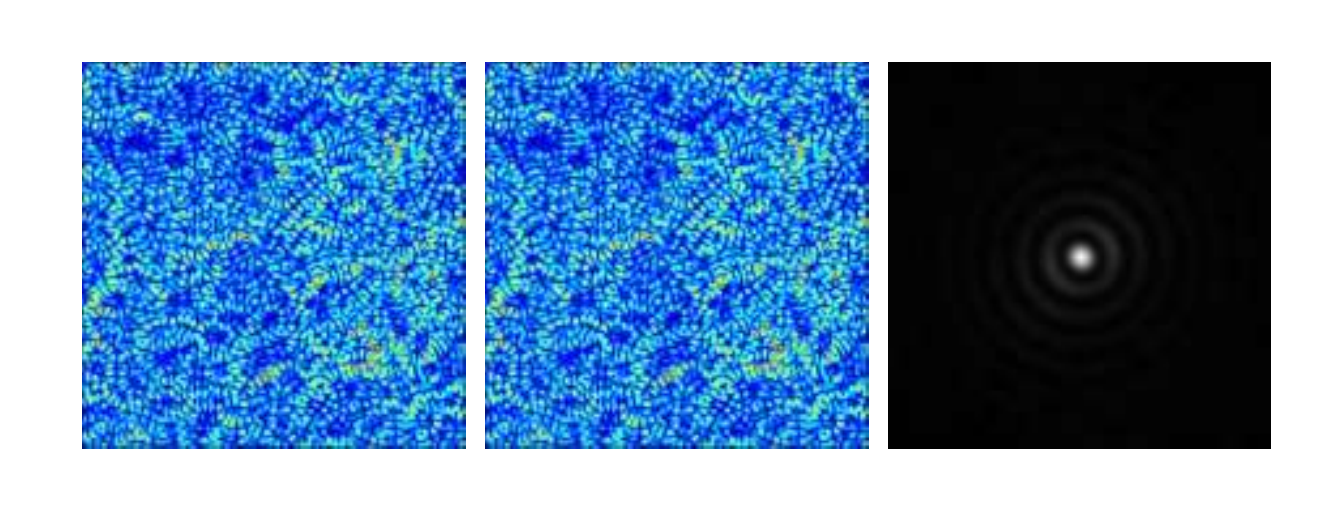}
    \caption{\textit{(Color online) The speckle patterns (left and middle) obtained by scattering a POV beam of order zero and the corresponding correlation function (right).}}
    \label{fig:speckle}
   \end{figure}
   \end{center}

Figure \ref{fig:speckle}, shows the speckle patterns obtained by the scattering of POV beam of order zero (left and middle) and the corresponding auto-correlation function (right). As discussed above, this correlation function is then used as the encryption key and the keys for decryption are generated as explained in Section 2. 

\section{Experiment and Simulation Results}

Our results are presented in this section to validate the proposed key generation for optical encryption system. We have used SRM University AP logo of 512 × 512 × 3 (RGB) and converted to gray scale with the size of 512 × 512 (see Fig. 4(a)) for easier computation. Fig. 4(b) shows the amplitude of the complex encrypted image using generated POV and as can be seen from the encrypted image, information is very difficult to be observed. Fig. 4(c) shows the decrypted image using appropriate secret keys. 

\begin{center}
	\begin{figure}[htb]
		\includegraphics[width=14.0cm]{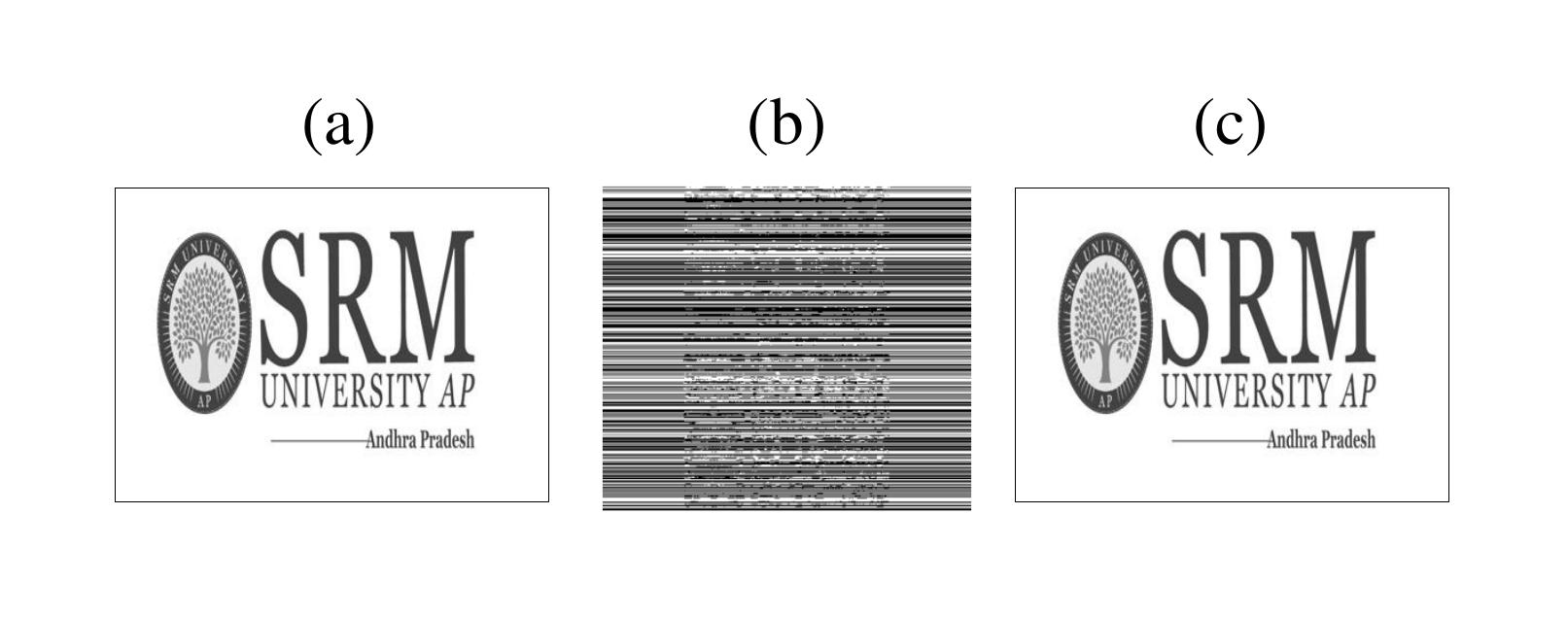}
		\caption{\textit{(Color online) Simulation results: (a) Input grayscale image, (b) encrypted image and (c) decrypted image (SSE = 2.7125e-21 dB).}}
		\label{fig:results}
	\end{figure}
\end{center}

To evaluate the decrypted image quality, we used sum squared error (SSE) metric which is calculated between the input image and decrypted image and is given as:

\begin{eqnarray}
SSE=10 log_{10}\left\lbrace \frac{\Sigma^{X}_{x=1}\Sigma^{Y}_{y=1}\lbrace \vert D(x,y)\vert -\vert E(x,y)\vert\rbrace^2}{\Sigma^{X}_{x=1}\Sigma^{Y}_{y=1}\lbrace \vert E(x,y)\vert\rbrace^2}\right\rbrace    
\end{eqnarray}

where D(x,y) represents decrypted image and E(x,y) denotes magnitude of an encrypted image \cite{attack,pcidrpe}. 

We recall that optical holographic imaging setup is preferred to record the encrypted complex image at the image plane (CCD). It is known that a precise optical alignment is required to record the encrypted hologram thus the image security is, in general, high \cite{inba_book}.   

\section{Conclusion}

In this work, we have proposed a protocol for secret key generation for optical encryption systems. We encoded the light information with the correlation function of generated POV speckles and found that this method further augments the security as ground glass plate scrambles the light field and makes it as a physically unclonable function. We further note that huge amount of information is needed to reconstruct the cross spectral density distribution, therefore, the complexity in reconstruction of the cross spectral density distribution enhances the security of the encryption protocol. Our results demonstrate that the intensity correlation function of POV speckles of different orders serve as a tool in optical information security.

\end{document}